\documentclass[prl,aps,showpacs,twocolumn]{revtex4}
\usepackage{amssymb}

\usepackage{graphicx}

\begin{document}

\title{DNA Spools under Tension}
\author{I. M. Kuli\'{c} and H. Schiessel}
\affiliation{Max-Planck-Institut f\"{u}r Polymerforschung, Theory
Group, POBox 3148, D 55021 Mainz, Germany}

\date{\today}

\begin{abstract}
DNA-spools, structures in which DNA is wrapped and helically coiled onto
itself or onto a protein core are ubiquitous in nature. We develop a general
theory describing the non-equilibrium behavior of DNA-spools under linear
tension. Two puzzling and seemingly unrelated recent experimental findings,
the sudden quantized unwrapping of nucleosomes and that of DNA\ toroidal
condensates under tension are theoretically explained and shown to be of the
same origin. The study provides new insights into nucleosome and chromatin
fiber stability and dynamics.
\end{abstract}


\maketitle

\textit{Introduction:} Wrapped DNA-protein complexes are ubiquitous in
nature~\cite{Saecker} and play key roles in many fundamental life processes.
Prominent examples of DNA wrapping proteins are: the Lac1 repressor~\cite
{Tsodikov} participating in the bacterial gene regulation, the DNA-gyrase~
\cite{Kampranis} directing changes in DNA topology, RNA polymerase~\cite
{Rivetti} copying DNA to RNA, and the histone octamer~\cite{luger97}
performing DNA packaging into nucleosomes leading in each cell to the
enormous condensation of meters of DNA into micron sized chromosomes.
Besides the natural wrapped architectures there are attempts to design
nanoparticles imitating that motive~\cite{Thurmond} as a means to
efficiently pack and transport DNA into cells. In most of these ligand-DNA
complexes the geometry and chemistry of the ligand surface enforces the DNA
to follow a superhelical wrapping path with one or more tight turns.
Remarkably, upon addition of multivalent condensing agents (like in sperm
cells) or under high crowding conditions (like in virus capsids or during $%
\psi $-condensation) DNA also shows an intrinsic ability to self-organize
into large toroidal spools~\cite{Bloomfield}.

In the past decade single molecule experiments have become
available allowing to apply tension to individual polymers in
order to probe their mechanical properties~\cite{Strick} as well
as their interaction with ligands~\cite{ligands,toroids,toroids2}
and molecular motors~\cite{motors}. Static and dynamic force
spectroscopy~\cite{DFS} developed into a powerful tool for
measuring equilibrium as well as kinetic characteristics of single
molecules, going far beyond the information accessible by
classical bulk experiments. Application of these methods to
DNA-spool geometries has been awaited for long and was reported
only recently for single nucleosomes~\cite {brower-toland02} and
single DNA\ toroidal condensates~\cite {toroids,toroids2}. These
experiments -- at first glance completely unrelated -- reside on
different length and energy scales and ground on different
mechanisms of wrapping. Despite that, they both reveal the same
surprising result apparently contradicting all the available bulk
data: the unfolding of wrapped DNA from the spools is a
catastrophic event, i.e., it is sudden and quantized and happens
one DNA turn at a time. The aim of this letter is to theoretically
explain this unusual nonequilibrium effect and to demonstrate the
universality behind it. Our theory is then applied to nucleosomes
and DNA toroids allowing us to extract from experiments the
relevant energetic parameters and to resolve apparent ''oddities''
in the dynamics of these systems.

\begin{figure}[tbp]
\includegraphics*[width=7.6cm]{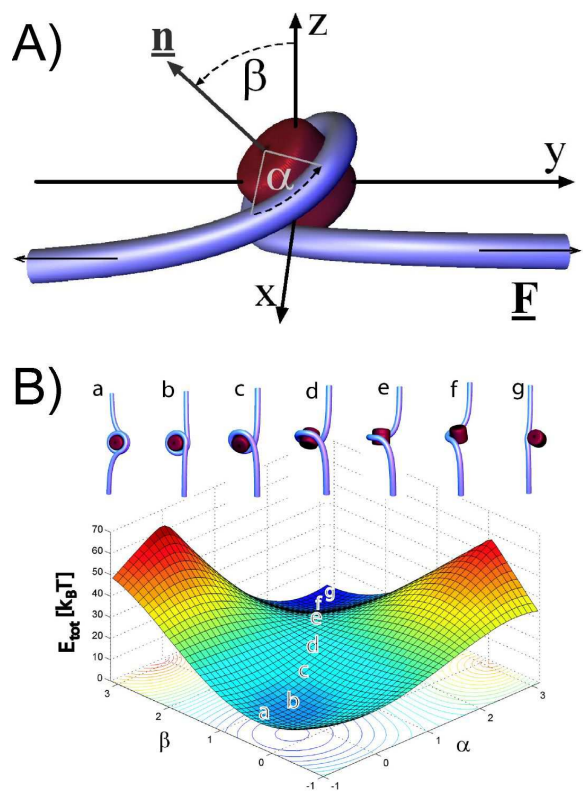}
\caption{A) An unfolding DNA-protein spool under tension is characterized by
two angles: the desorption angle $\protect\alpha $ and the tilt angle $%
\protect\beta $. B) The energy landscape of a DNA-spool (as given
by Eq.~\ref {Etot} for $A=50nm$, $F=4pN,$ $\protect\varepsilon
_{ads}=0.7k_{B}T/nm$, $R=4.2nm$, $H=2.5nm)$ and its unfolding
pathway.} \label{Tsodikov}
\end{figure}

\textit{General Model:} The DNA is assumed to be adsorbed on the protein
spool surface along a predefined helical path with radius $R$ and pitch
height $H$. This path accounts for the typical chemical structure of such a
protein spool surface (e.g. distribution of charges, hydrogen
donors/acceptors etc.) along which the DNA is adsorbed with a net adsorption
energy density $\varepsilon _{ads}$ given by the difference of the total
DNA-protein binding energy and the stored DNA bending energy per length
along the helical path. The degree of DNA adsorption is described by the
desorption angle $\alpha $ which is defined to be zero for one full turn
wrapped (cf. Fig.~1). After short inspection it becomes clear that the
unwrapping problem is non-planar and that the spool needs to rotate
transiently out of the plane while performing a full turn -- an effect
already pointed out by Cui and Bustamante~\cite{cui00}. Therefore a second
angle, $\beta $, is introduced to describe the out-of-plane tilting of the
spool. When a tension $F$ (along the $Y$--axis) acts on the two outgoing DNA
''arms'' the system (i.e., the wrapped spool together with the free DNA
ends) will simultaneously respond with (\textit{i}) \textit{DNA deformation}%
, with (\textit{ii}) \textit{spool tilting} and with (\textit{iii}) \textit{%
DNA desorption} from the spool. In the following we assume that the DNA has
freely rotating ends (as in the experiment~\cite{brower-toland02}) that allows
us to neglect the twist degree of freedom. Then the total
energy of the system as a function of $\alpha $ and $\beta $ writes $%
E_{tot}=2R\varepsilon _{ads}\alpha +2E_{bend}-2F\Delta y$. Here the first
term is the adsorption energy, the second the bending deformation energy of
the two free DNA portions, and the third term describes the gain in
potential energy by pulling out the DNA ends by a distance $2\Delta y$.
Applying linear elasticity theory and elementary geometry $E_{tot}$ can be
redistributed into the following three terms:
\begin{equation}
E_{tot}\left( \alpha ,\beta \right) =E_{comp}\left( \alpha \right)
+E_{geom}\left( \alpha ,\beta \right) +E_{stiff}\left( \alpha ,\beta \right)
\label{Etot}
\end{equation}
The first term $E_{comp}=2R\left( \varepsilon _{ads}-F\right) \alpha $
describes the competition of the adsorption and the applied force. The
''geometrical'' energy term $E_{geom}=2FR\left( \cos \beta \sin \alpha
-\left( H/2\pi R\right) \left( \pi -\alpha \right) \sin \beta \right) $
stems from the gain/loss of potential energy by spool opening (change of $%
\alpha $) and rotation (change of $\beta $). Finally, the last and most
remarkable term $E_{stiff}=8\sqrt{AF}\allowbreak \left( 1-\sqrt{\left(
1+\left( R/\overline{R}\right) \cos \beta \cos \alpha +\left( H/2\pi
\overline{R}\right) \sin \beta \right) /2}\right) $ accounts for the
stiffness of the non-adsorbed DNA portions. Here $A\approx 50k_{B}Tnm$ is
the DNA bending stiffness at room temperature and $\overline{R}%
^{2}=R^{2}+H^{2}/4\pi ^{2}$. Two effects contribute equally to $E_{stiff}$:\
1) the bending energy of the deformed DNA arms and 2) the loss of potential
energy by ''wasting'' length due to DNA deformation. To understand the
implications of Eq.~\ref{Etot} on the kinetics of unwrapping we consider two
limiting cases. First let us look at the case of a large thin spool, i.e., $%
R\gg A/k_{B}T$ (or, equivalently, an infinitely flexible polymer)
and $R\gg H $, where we may neglect $E_{stiff}$. In that case and
for $F>\varepsilon _{ads}$ the spool moves from the
(thermodynamically) metastable state $M_{1}$ with $\alpha =\alpha
_{0}=-\arccos \left( 1-\varepsilon _{ads}/F\right) $ and $\beta
=0$ via a saddle point $S$ at $\alpha =0$ and $\beta =\alpha _{0}$
into a more favorable minimum $M_{2}$ at $\alpha =\pi +\alpha _{0}$ and $%
\beta =\pi $. Remarkably $S$ constitutes a significant energetic barrier
between $M_{1}$ and $M_{2}$ given by $\Delta E_{tot}=2FR\left( \alpha
_{0}\cos \alpha _{0}-\sin \alpha _{0}\right) $. For hypothetical yet
reasonable parameter values, say $R=50nm, $ $\varepsilon _{ads}=1k_{B}T/nm$
and $F=2\varepsilon _{ads}$ we obtain a huge barrier of $\Delta
E_{tot}\approx 70k_{B}T$! A second interesting limit of Eq.~\ref{Etot} is
given by a flat spool and high polymer stiffness, i.e., $A\gg Rk_{B}T$ and $%
R\gg H$. For not too large forces ($F\lesssim A/R^{2}$) and $\varepsilon
_{ads}\lesssim F$ the kinetic behavior is roughly dominated by the term $%
E_{stiff}$. In this case we find a transition path from $(\alpha ,\beta
)=(0,0)$ over the saddle point $(\pi /2,\pi /2)$ to the state $(\pi ,\pi )$
with a barrier height $\Delta E_{tot}=8\sqrt{AF}\left( 1-1/\sqrt{2}\right) $%
. Note that in this limit the DNA actively participates in the suppression
of unwrapping ($\Delta E_{tot}\sim A^{1/2}F^{1/2}$) which can even give rise
to negative resistance effects~\cite{negative resistance} for small forces.
In preliminary conclusion, in both limiting cases the unwrapping meets
significant kinetic barriers but for different reasons: because of
unfavorable projection of the force in terms of the $\left( \alpha ,\beta
\right) $ configurational space in the first limit and due to significant
transient bending of the DNA arms during the transition in the second limit.
For realistic DNA-spools we are somewhere in between these two cases.

\textit{Nucleosome Unwrapping}: The most abundant DNA spool in nature is the
nucleosome where 1 and $3/4$ turns of DNA, 147 bp, are wrapped around a protein core
on a lefthanded superhelical path with diameter $4.2nm$ and $2.5nm$ pitch.
The question about the equilibrium and
kinetic stability of nucleosomes is one of the important experimentally
unsettled questions in present molecular biology. How can nucleosomes be
highly stable with its wrapped DNA being highly accessible at the same time~
\cite{polach95}? A recently performed experiment~\cite{brower-toland02}
measuring the critical force required to unwrap single nucleosomes reveals
an interesting and unexpected behavior~\cite{Foot4}. When small forces ($F<10pN$) are
applied for short times ($\sim 1-10$ $s$) the nucleosome unwraps only
partially by releasing the outer 60-70 bp of wrapped DNA (moving from state
a to b in Fig.~1B) in a gradual and equilibrium fashion. For higher forces ($%
F\gtrapprox 20pN$) nucleosomes show a pronounced sudden
non-equilibrium release behavior of the remaining 80 bp (cf. c-g
in Fig.~1B) -- the latter force being much larger than expected
from equilibrium arguments~\cite{markonetz}. In fact, experiments~\cite{polach95}
measuring spontaneous partial unwrapping of nucleosomal DNA suggest $30k_{B}T$ per
147bp leading to an unpeeling force of $\sim 2.5pN$.
To explain this
peculiar finding Brower-Toland et al.~\cite {brower-toland02}
conjectured that there must be a barrier in the adsorption energy
located after the first 70-80 bp which reflects some biochemical
specificity of the nucleosome structure at that position. Their
analysis of the dynamical force spectroscopy measurements revealed
an apparent barrier of $\sim 38k_{B}T$ smeared out over not more
than 10 bp. However, there is no experimental indication of such a
huge specific barrier -- neither from the crystal
structure~\cite{luger97} nor from the equilibrium accessibility to
nucleosomal DNA~\cite{polach95}. Consequently the question arises
if the barrier is really caused by biochemistry of the nucleosome
or, as we show below, by its underlying geometry and physics.

\begin{figure}[tbp]
A) \includegraphics*[width=6.5cm]{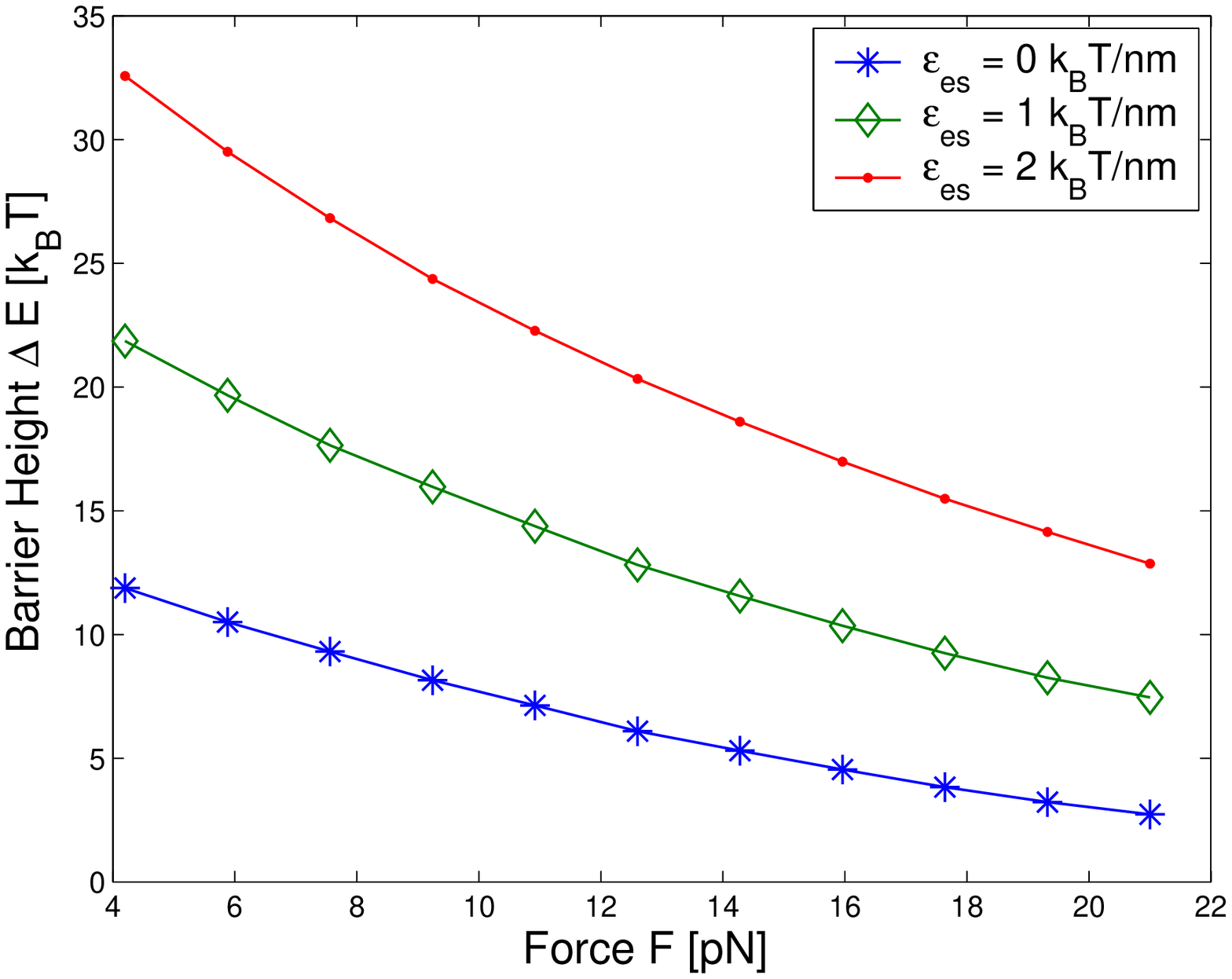} \newline
B) \includegraphics*[width=6.5cm]{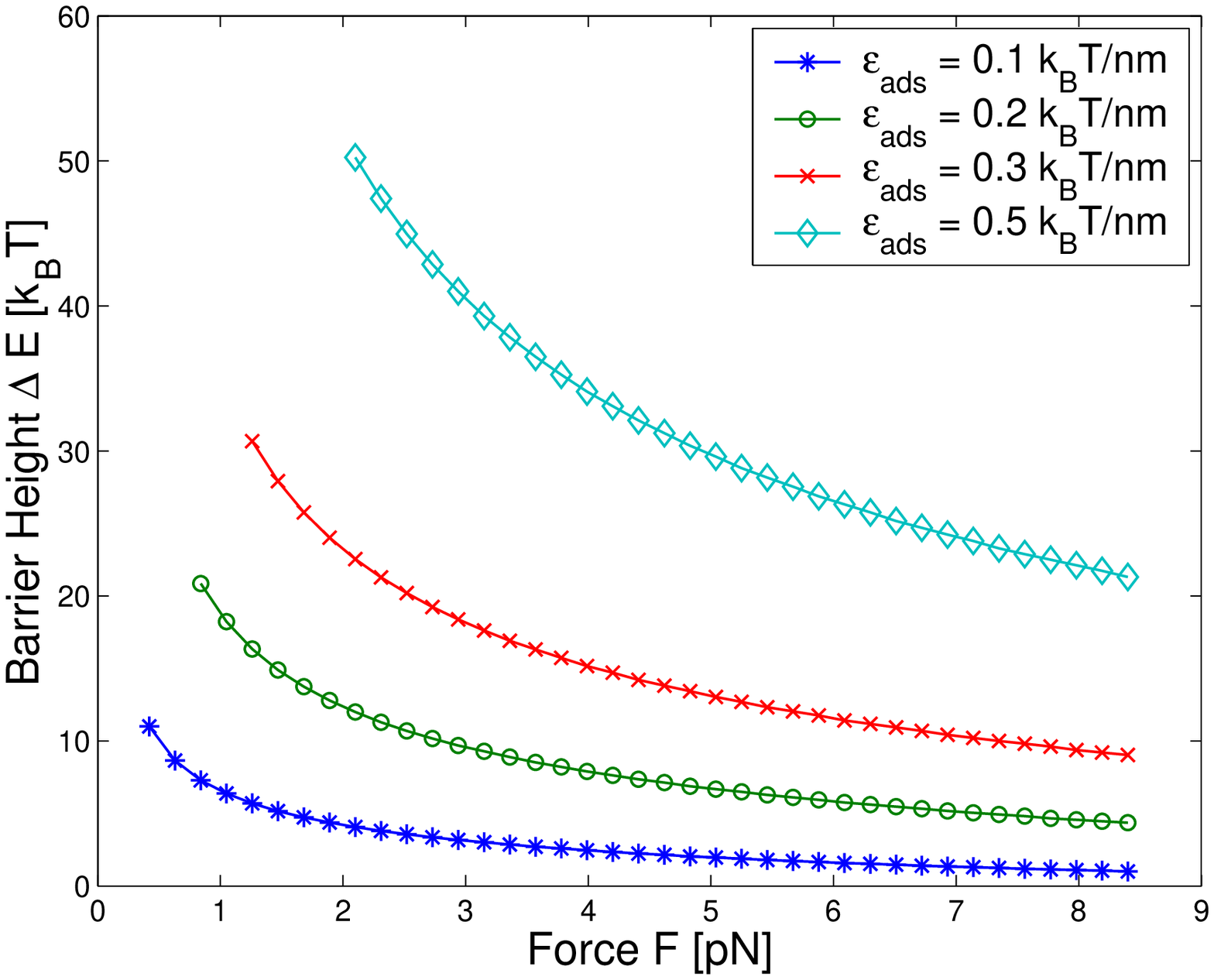}
\caption{Kinetic barriers opposing DNA-spool unfolding as function of
applied tension computed for: A) The nucleosome ($R=4.2nm,$ $H=2.4nm,$ $%
\protect\varepsilon _{ads}^{0}=0.7k_{B}T/nm,$ $A=50k_{B}Tnm,$cf. text) for
various interstrand repulsion energy densities $\protect\varepsilon _{es}$
and B) the DNA toroid ($R=50nm,$ $H=2.4nm,$ $A=40k_{B}Tnm$) for various
adsorption energy densities $\protect\varepsilon _{ads}$. }
\end{figure}

\begin{figure}[tbp]
\includegraphics*[width=7cm]{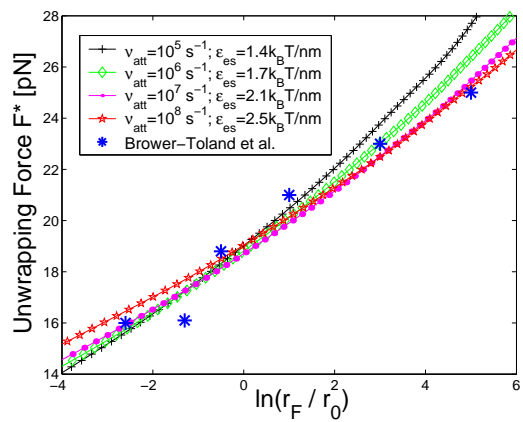}
\caption{Optimal fits of the DFS data from Ref.~\cite{brower-toland02}
for various attempt frequencies give the corresponding electrostatic
DNA-DNA-repulsion $\protect\varepsilon _{es}$. }
\end{figure}

To see that the effect is mainly physical we apply Eq.~\ref{Etot}
to compute the barrier. For this purpose we model the nucleosomal
adsorption energy density as $\varepsilon _{ads}\left( \alpha
\right) =\varepsilon
_{ads}^{0}+\theta \left( \alpha \right) \varepsilon _{es}$ where $%
\varepsilon _{ads}^{0}\approx 0.7k_{B}T/nm$ is taken from the reversible
part (for the first 60-70bp) of the measurement in Ref.~\cite
{brower-toland02}. The introduction of the step function ($\theta =0$ for $%
\alpha <0$ and $\theta =1$ for $\alpha \geq 0$) together with a
new parameter $\varepsilon _{es}$, the electrostatic interaction
energy density, accounts for the DNA-DNA repulsion of the two
adjacent helical gyres which acts only for $\alpha <0$ reducing
the net $\varepsilon _{ads}$ (cf. Fig.~1A). Using $\varepsilon
_{ads}^{0}$ from above we can compute the barrier height for nucleosome
unfolding for various values of $\varepsilon _{es}$ as done in
Fig.~2A. To relate the barrier heights from Fig.~2A to the
dynamical force spectroscopy (DFS) measurements in Ref.
\cite{brower-toland02} we generalize the classical relation
between the loading rate $r_{F}$ and the most probable rupture
force $F^{\ast }$ \cite{DFS} to the case of nonlinear
force-barrier dependence and obtain $\ln \left( r_{F}/r_{0}\right)
=\ln \left[ -\nu _{att}k_{B}T/\left( r_{0}\Delta E^{\prime
}\right) \right] -\Delta E/k_{B}T.$ Here $r_{F}$ and $\Delta E$
are functions of $F^{\ast }$ and $\Delta
E^{\prime }=\partial \left( \Delta E\right) /\partial F^{\ast }$. $%
r_{0}=1pNs^{-1}$ is an arbitrary scale on the $r_{F}$ axis and $\nu _{att}$
is the typical attempt frequency of the nucleosome. Assuming $\nu _{att}$ in
the range $10^{5}$ to $10^{8}s^{-1}$ we can fit the experimental data from
\cite{brower-toland02} to obtain the corresponding values of $\varepsilon
_{es}$, cf. Fig.~3. Keeping in mind that $\nu _{att}$ is dominated by the slowest
process involved in the unfolding event we estimate $\nu
_{att}\lessapprox 10^{6}s^{-1}$ \cite{Foot3}. The latter implies $\varepsilon
_{es}\approx 1.4-1.7k_{B}T/nm.$ So indeed, at the line $\alpha =0$ there is
clear jump in adsorption energy density as we would naively expect from
repulsive DNA-DNA electrostatics under these conditions~\cite{Foot1.1}. This
explains why under equilibrium conditions (at $F=0$) the DNA deeply inside
the nucleosomes (almost the whole bound DNA!) can be rather easily accessed
by proteins~\cite{polach95} but the nucleosome is still highly stable: The
line $\alpha =0$ can be moved to each position inside the nucleosome if the
left and right DNA arms are adsorbed/desorbed in a consistent manner. Beyond
that ($\alpha >0$) the assisting electrostatics switches off and the
nucleosome is suddenly strongly stabilized (by $60-70k_{B}T$ in total!).

\textit{Toroid Unspooling:} When long DNA molecules condensed with
multivalent counterions were stretched in a single molecule
experiment~\cite {toroids} Baumann et al. found a surprising
behavior. When a critical force (typically $F\approx 4-12pN$) is
reached large portions of DNA are released in packets in a
discontinuous manner (''stick release pattern''). When the same
experiment was redone recently by another group~\cite{toroids2} a
pronounced quantization in the DNA release length of $\approx
300nm$ was clearly demonstrated. It was noted in
Refs.~\cite{toroids,toroids2} that the latter correlates exactly with
the typical size ($R\approx 50nm$) of toroidal condensates formed
in solution and led those authors to the conclusion that a single
turn of DNA unwraps from the toroid spool at a time. Despite that
interesting finding the mechanism behind this non-equilibrium
effect remained unexplained. However in the light of our theory the
explanation is again straightforward as a DNA-toroid exhibits a
spool geometry with $R\approx 50nm$ and $H\ll R$. The
''first limit'' considered above gives here a good approximation. The
barrier heights for different values of $\varepsilon _{ads}$ as a function of force
are computed in Fig.~2B. Similar as in the case of the nucleosome the attempt
frequency $\nu _{att}$ is dominated by the rotational friction --
here of the $50nm$ sized toroid object -- leading to
$\nu_{att}\approx 3\times \left( 10^{2}-10^{3}\right) s^{-1}$. For high
concentrations of the condensing agent spermidine one finds
$\varepsilon _{ads}\approx 0.2-0.3k_{B}T/nm$ (cf.
\cite{toroids,toroids2} and the references therein). In case of equilibrium
this means a very small peeling-off force of
$F\approx $ $1-1.5pN$. Our model together with Fig.~2B allows us now to predict the
activated non-equilibrium behavior to have very low unfolding
frequencies $\nu
_{unf}=v_{att}\exp \left( -\Delta E\left( F\right) /k_{B}T\right) $, for instance $%
10^{-6}$ to $10^{-2}s^{-1}$ for $F=2pN$ , $10^{-3}$ to $1s^{-1}$ for $F=4pN$
and $0.3$ to $50$ $s^{-1}$ for $F=8pN$~\cite{Foot2} -- consistent
with experimental findings~\cite{toroids,toroids2}.

\textit{Conclusion: }We have shown that DNA-spools ranging from protein-DNA
complexes to DNA toroids share a universal feature inherited by their
geometry: They are strongly kinetically protected from mechanical disruption
upon applied tension. In the case of chromatin fibers consisting of large
arrays of nucleosomes and other DNA spooling proteins this effect provides a
great biological advantage. Strong molecular motors like RNA polymerase and
helicase or microtubuli during cell division are known to act on the fiber
with significant transient tensions of the order of$\ 20pN$ or even more.
While a hypothetical ''fiber A'' consisting of DNA and non-spooling proteins
(say only DNA bending proteins) would immediately lose most of its protein
content a ''fiber B'' constituted of DNA-spools would survive long time
periods (up to $10^{6}-10^{8}$ times longer than ''fiber A''). We can
speculate that this obvious advantage was not overlooked by nature and has
flown into the chromatin fiber design and the nucleosome-spool shape. The
remarkable universality of the ''kinetic protection'' also shows up in the
case of DNA-toroids which are roughly $\sim 10$ times larger while the DNA
is $\sim 10$ times weaker adsorbed than for typical DNA-protein spools.
While the biological implications of this finding still have to be fully
figured out it seems that this might play a role in the injection / ejection
process of DNA from viral capsids inducing similar quantization effects as
found here. Looking at the wealth of peculiar effects revealed by the single
molecule experiments~\cite{toroids,toroids2} we feel that the present
understanding of DNA condensation kinetics is still incomplete, yet one
partial mystery seems resolved.


\begin{thebibliography}{99}
\bibitem{Saecker}  Reviewed in R.M Saecker and M.T. Record, Curr. Op.
Struct. Biol. \textbf{12}, 311 (2002)

\bibitem{Tsodikov}  O.V. Tsodikov et al., J. Mol. Biol.\ \textbf{294}, 639
(1999)

\bibitem{Kampranis}  S.C. Kampranis, A.d. Bates and A. Maxwell, Proc. Natl.
Acad . Sci. USA\textbf{\ 96}, 8414 (1999)

\bibitem{Rivetti}  C. Rivetti, M. Guthold and C. Bustamante, EMBO J. \textbf{%
18}, 4464 (1999)

\bibitem{luger97}  K. Luger, A. W. M\"{a}der, R. K. Richmond, D. F. Sargent,
and T. J. Richmond, Nature (London) \textbf{389}, 251 (1997).

\bibitem{Thurmond}  K.B. Thurmond, E.E. Remsen, T. Kowalewski and K.L.
Wooley, Nucl. Acids Res. \textbf{27}, 2966 (1999)

\bibitem{Bloomfield}  Reviewed in V.A. Bloomfield, Curr. Op. Struct. Biol.
\textbf{6}, 334 (1996); Biopolymers. \textbf{44}: 269 (1997).

\bibitem{Strick}  T. R. Strick et al. Rep. Prog. Phys. \textbf{66}, 1 (2003)

\bibitem{ligands}  M.C. Wiliams et al. Proc. Natl. Acad. Sci. USA \textbf{98}%
, 6121 (2001); M. Hegner, S. B. Smith and C. Bustamante, Proc. Natl. Acad.
Sci. USA \textbf{96}, 10109 (1999);

\bibitem{toroids}  C. G. Baumann et al. Biophys. J. \textbf{78}, 1965 (2000);

\bibitem{toroids2}  Y. Murayama, Y. Sakamaki and M. Sano. Phys. Rev. Lett.
\textbf{90}, 018102 (2003)

\bibitem{motors}  K. Svoboda and S.M. Block Cell \textbf{77}, 773 (1994); H.
Yin et al. Science \textbf{270}, 1653 (1995)

\bibitem{DFS}  E. Evans and K. Ritchie, Biophys J.\textbf{\ 72}, 1541
(1997); E. Evans Biophys. Chem. \textbf{82, }83 (1999)

\bibitem{brower-toland02}  B. D. Brower-Toland et al. Proc. Natl. Acad. Sci.
USA \textbf{99}, 1960 (2002).

\bibitem{cui00}  Y. Cui and C. Bustamante Proc. Natl. Acad. Sci. USA \textbf{%
97}, 127 (2002).

\bibitem{negative resistance}  G.C. Cecchi and M.O. Magnasco, Phys. Rev.
Lett. \textbf{76}, 1968 (1996); D. Bartolo, I. Der\'{e}nyi and A. Ajdari,
Phys. Rev. E, \textbf{65}, 051910 (2002)

\bibitem{polach95}  K. J. Polach and J. Widom, J. Mol. Biol. \textbf{254},
130 (1995); \textbf{258}, 800 (1996).

\bibitem{Foot4}  The experiments were performed on DNA chains with up to
17 nucleosomes complexed at well-defined positions. In the force range
of interest their coupling can be savely neglected
since the intranucleosomal distance $d\sim 40nm$
exceeds the DNA-linker induced interaction length $\lambda \sim \sqrt{A/F}$,
cf. R. Bruinsma and J. Rudnick, Biophys. J. \textbf{76}, 1725 (1999).

\bibitem{markonetz}  J. F. Marko and E. D. Siggia, Biophys. J. \textbf{73},
2173 (1997); K.-K. Kunze and R. R. Netz, Phys. Rev. E \textbf{66}, 011918
(2002)

\bibitem{Foot3}  The rotational attempt frequency of a nucleosome-sized
sphere is of the order $2\frac{k_{B}T}{8\pi R^{3}\eta _{s}}%
\approx 10^{5}-10^{6}s^{-1}$ with $\eta_s$ being the water viscosity
(a centipoise). The typical frequency that
characterizes the relaxation of the DNA arms is comparable to that -- even
if one accounts for additional complexed nuclesomes as it is the case in the
experiment~\cite{brower-toland02}. There the first unfolding nucleosome is
surrounded by 16 other nucleosomes that have to move via a distance $\Delta
s\approx 25nm$ (unfolding length) under a force of $F\approx 10-20pN$
leading to an lower bound $10^{5}-10^{6}s^{-1}$ of that frequency.

\bibitem{Foot1.1}  Note that the latter alone would not explain the sudden
catastrophic behaviour and the slow kinetics of unfolding if the effects
described by Eq.\ref{Etot} were not included explicitly.

\bibitem{Foot2}  The upper/lower estimate correspond to $\varepsilon
_{ads}=0.2$ / $0.3k_{B}T/nm$; note the strong dependence of $\nu _{unf}$ on $%
\varepsilon _{ads}.$
\end{thebibliography}
\end{document}